\documentstyle[preprint,aps,prd]{revtex}

\tighten
\begin{document}

\draft

\title{Numerical Evolution of Dynamic 3D Black Holes:  Extracting Waves}

\author{Karen Camarda${}^{(1,4,5)}$ and Edward Seidel${}^{(1,2,3,4)}$}
\address{
${}^{(1)}$ Max-Planck-Institut f{\"u}r Gravitationsphysik,
Schlaatzweg 1, 14473 Potsdam, Germany
}
\address{
${}^{(2)}$ National Center for Supercomputing Applications,
Beckman Institute, 405 N. Mathews Ave., Urbana, IL, 61801
}
\address{
${}^{(3)}$ Department of Astronomy,
University of Illinois, Urbana, IL 61801
}
\address{
${}^{(4)}$ Department of Physics,
University of Illinois, Urbana, IL 61801
}
\address{
${}^{(5)}$ Department of Astronomy and Astrophysics and Center for
Gravitational Physics and Geometry, \\
Pennsylvania State University, University Park, PA 16802
}
\maketitle

\begin{abstract}
We consider the numerical evolution of dynamic black hole initial data
sets with a full 3D, nonlinear evolution code.  These data sets
consist of single black holes distorted by strong gravitational waves,
and mimic the late stages of coalescing black holes.  Through
comparison with results from well established axisymmetric codes, we show that
these dynamic black holes can be accurately evolved.  In particular,
we show that with present computational resources and techniques, the
process of excitation and ringdown of the black hole can be
evolved, and one can now extract accurately the gravitational waves
emitted from the 3D Cartesian metric functions, even though they may be
buried in the metric at levels on the order of $10^{-3}$ and below.
Waveforms for both
the $\ell=2$ and the much more difficult $\ell=4$ modes are computed
and compared with axisymmetric calculations.  In addition to exploring
the physics of distorted black hole data sets, and showing the extent
to which the waves can be accurately extracted, these results also
provide important testbeds for all fully nonlinear numerical codes
designed to evolve black hole spacetimes in 3D, whether they use
singularity avoiding slicings, apparent horizon boundary conditions,
or other evolution methods.

\end{abstract}

\section{Introduction}  As numerical relativity is empowered by ever
larger computers,
numerical evolutions of black hole data sets are becoming more and
more common\cite{Seidel96b}.  The need for such simulations is great,
especially as gravitational wave observatories are gearing up to collect
gravitational
wave data over the next decade\cite{Abramovici92}.  As black hole
collisions are considered a most promising source of signals to be
detected by these observatories, it is crucial to have a detailed
theoretical understanding of the coalescence process that can only be
achieved through numerical simulation.  In particular, it is most
important to be able to simulate accurately the excitation of the
coalescing black holes, to follow the waves generated in the process,
and to extract gravitational waveforms expected to be seen by
detectors.

This is a very difficult calculation, as one must simultaneously deal
with singularities inside the black holes, follow the highly nonlinear
regime in the coalescence process taking place near the horizons, and
also calculate the linear regime in the radiation zone where the
waves represent a very small perturbation on the background spacetime
metric.  In axisymmetry this has been achieved for distorted black
holes with rotation \cite{Brandt94c} and without \cite{Abrahams92a},
and for equal mass colliding black holes \cite{Anninos93b,Anninos94b},
but with difficulty.  These 2D evolutions can be carried out to roughly
$t=100M$, where $M$ is the ADM mass of the spacetime, although beyond
this point large gradients related to singularity avoiding slicings
usually cause the codes to become very inaccurate and crash.

In such simulations, it has been shown that, using a gauge-invariant
radiation extraction technique developed originally by
Abrahams\cite{Abrahams88,Abrahams90}, one can in principle extract the
waveforms
generated by the black holes.  However, even in axisymmetric simulations,
where the
coordinate systems are naturally adapted to the black holes and the
radiation, these waveforms can sometimes be difficult to compute cleanly.
The waves are small perturbations buried in the metric functions
actually being evolved, are generated in the strong field regime just
outside the evolving horizon, and then propagate out to the wave zone
where they must be extracted.  The energy carried by these waves is
typically found to be on the order of $10^{-6}-10^{-2}M$.
At such low amplitude, both the generation and
propagation of these signals are susceptible to small numerical errors
inherent in numerical simulations.  For example, in the axisymmetric
evolution of Misner data for two colliding black holes, although the
$\ell=2$ signals have been accurately computed, as verified by careful
comparisons with perturbation theory in different regimes, the more
difficult $\ell=4$ signals are still rather
uncertain\cite{Anninos94b}.

In this paper we show that with current techniques and computational
resources available to 3D numerical relativity, distorted black holes
can be evolved through the initial relaxation and final ringdown
period, and that the gravitational waveforms can be followed and
accurately extracted from the numerical evolutions, even though they
represent a small perturbation on the background spacetime which is
also being evolved.  Here we focus on the application of the 3D code
to the evolution of an axisymmetric distorted black hole initial
data set, so that careful comparisons can be made with results obtained
with mature 2D codes.  The evolutions can be carried out at present
through about $t=30-35M$, which provides time to study several
wavelengths of the fundamental $\ell=2$ and $\ell = 4$ modes present
in the simulations.  The extension to full 3D black hole initial data and wave
modes, for which no testbeds exist at present, as well as more
extensive comparisons with axisymmetric initial data, are in progress
and will be published elsewhere.  In particular, {\em nonaxisymmetric}
modes, such as
the $\ell=2,m=2$ mode expected to be important in the ringing radiation for
rotating
black holes at late times~\cite{Flanagan97a}, and therefore an important
signal for gravitational wave observations,
can now be studied\cite{Camarda97a}.

\section{3D Evolution of Black Holes with Radiation}
\subsection{Our code and prior 3D simulations}
We have
developed a 3D code to study black holes and gravitational waves in
Cartesian coordinates.  This code (known as the ``G'' code) was
applied to Schwarzschild black holes, where we showed that using
singularity avoiding time slicings, a spherical black hole could be
evolved accurately to $t=30-50M$, depending on the resolution, location of the
outer boundary, and the slicing conditions\cite{Anninos94c}.  Beyond
that time, the code generally crashes due to the unbounded growth of
metric functions generated by singularity avoiding slicings.
However, the focus of\cite{Anninos94c}
was on spherical black holes, so no studies were made of black hole
oscillations and the waves that would be generated in the process.  It was
shown that with spherical initial data some nonspherical behavior
could be introduced by the Cartesian mesh and boundary conditions, the
numerics of which could in principle generate spurious gravitational
waves.

The same code was simultaneously applied to the problem of pure
gravitational waves\cite{Anninos94d}, where many systems were studied,
from pure linear quadrupole waves to nonlinear waves, and their
propagation on a Cartesian mesh was studied.  In that study it was
shown that waves can be accurately evolved, although certain problems
with gauge modes in the ``near linear'' regime that can confuse the
results were identified, along with strategies to deal with them.

This 3D code was then applied to the collision of two axisymmetric
black holes (Misner data) \cite{Anninos96c}, where we showed by
comparison to 2D results that one could accurately track the merging
of the horizons, and that the radiation emitted was qualitatively the
same, but at that time the waveforms were not studied extensively.
Building on the work presented in this paper, a more detailed study of
the Misner data in 3D, including the waveforms, is in preparation for
publication
elsewhere.

In this paper, we build on this prior work, focusing on the 3D evolution of
distorted single black
holes.  In previous 2D axisymmetric studies, such data sets were
shown to be similar to single black holes formed just after the merger
of colliding axisymmetric black holes.  The agreement between the waveforms
produced with the 3D and 2D codes is the subject of this paper, while
comparisons of metric functions and other quantities for a variety
of axisymmetric initial data sets will be found
in~\cite{Camarda97a,Camarda97c}.

\subsection{Initial data} The initial data we evolve in this paper
consist of a single black hole that has been distorted by the
presence of an adjustable torus of nonlinear gravitational waves which
surround it.  The amplitude and shape of the torus can be specified by hand,
as described below, and can create very highly distorted black holes.
Such initial data sets, and their evolutions in axisymmetry, have been
studied extensively, as described in
Refs.\cite{Abrahams92a,Bernstein94a,Bernstein93b}.  For our purposes,
we consider them as convenient initial data that create a distorted
black hole that mimics the merger, just after coalescence, of two
black holes colliding in axisymmetry \cite{Anninos94b}.

Following\cite{Bernstein94a}, we write the 3--metric
as
\begin{equation}
d\ell^2 = \tilde{\psi}^4 \left( e^{2q} \left( d\eta^2 + d\theta^2 \right) +
  \sin^2\theta d\phi^2 \right),
\end{equation}
where $\eta$ is a radial coordinate related to the Cartesian
coordinates by $\sqrt{x^{2}+y^{2}+z^{2}} = e^{\eta}$. (We have set
the scale parameter M in\cite{Bernstein94a} to be 2 in this paper.) Given a
choice for the ``Brill wave'' function $q$, the Hamiltonian constraint
leads to an elliptic equation for the conformal factor $\tilde{\psi}$.  The
function $q$ represents the gravitational wave surrounding the black
hole, and is chosen to be
\begin{equation}
\label{eq:q2d}
q\left(\eta,\theta,\phi\right) = a \sin^n\theta \left(
  e^{-\left(\frac{\eta+b}{w}\right)^2} +
  e^{-\left(\frac{\eta-b}{w}\right)^2} \right)
  \left(1+c \cos^2\phi\right).
\end{equation}
Thus, an initial data set is characterized by the parameters
$\left(a,b,w,n,c\right)$, where, roughly speaking, $a$ is the amplitude
of the Brill wave, $b$ is its radial location, $w$ its width, and $n$
and $c$ control its angular structure.  Note that we have generalized
the original axisymmetric construction to full 3D by the addition of
the parameter $c$, but in this paper we restrict ourselves to $c=0$
for comparison with axisymmetric results.  A study of full 3D initial
data and their evolutions will be published elsewhere
\cite{Camarda97a,Camarda97c,Brandt97a}.  If the amplitude $a$ vanishes, the
undistorted
Schwarzschild solution results, leading to
\begin{equation}
\tilde{\psi} = 2 \cosh \left( \frac{\eta}{2} \right).
\end{equation}

We note that just as the Schwarzschild geometry
has an isometry that leaves the metric unchanged under the operation
$\eta \rightarrow -\eta$, our data sets also have this property, even in the
presence of the Brill wave.  As discussed in
\cite{Anninos94c,Bernstein93b}, this condition can also be applied during
the evolution and in Cartesian coordinates as well.  The evolution of the
data set
$(0.5,0,1,2,0)$ is considered in this paper.  In what follows we solve
the Hamiltonian constraint for this initial data set, interpolate
it onto a 3D Cartesian grid, and study its evolution with a 3D
evolution code.

\subsection{Evolution and Analysis} Using the techniques described
in\cite{Anninos94c},
we evolve the initial data sets described above in
3D Cartesian coordinates.  The present evolution code is based on the
one detailed in\cite{Anninos94c,Anninos94d}, using the same finite
difference
algorithms, having the same convergence properties, {\it etc.}, but having
been rewritten to take advantage of newer parallel computers.

Although the 3D evolution code is written without making use of any
symmetry assumptions, the initial data we evolve in this paper have
both equatorial plane symmetry and axisymmetry.  Hence we save on
the memory and computation required by evolving only one octant of
the system.  As shown in \cite{Anninos94c}, this has no effect on the
simulations except to
reduce the computational requirements by a factor of eight.  Even with such
computational savings, these are extravagant calculations.  The
results presented in this paper were computed on a 3D Cartesian grid of
$300^{3}$ numerical grid zones, which is about a factor three larger than
the largest
production relativity calculations of which we are aware (which were
about $200^{3}$ zones).  With our new
code, these take about 12 Gbytes of memory, and require about a day
on a 128 processor, early access SGI/Cray Origin 2000 supercomputer.

Given a choice of lapse and shift, the Cartesian metric functions
$\gamma_{xx}, \gamma_{xy}$, {\it etc.}, are evolved using the ADM formulation
of the Einstein equations.  In this paper we use a lapse which is
initially maximal, with antisymmetric conditions across the throat of
the black hole, defined by the isometry surface $\eta=0$, or $r=1$.
The initial data are then evolved with the ``1+log'' algebraic lapse
condition\cite{Anninos94c}, an isometry operator in Cartesian
coordinates, and with
zero shift.  These choices have been made for computational
efficiency, and are not unique choices for successful evolution.  For
example, we have performed similar calculations with maximal slicing
and no isometry with similar results, except that the computational
time needed to solve the elliptic maximal slicing equation can double
or triple the computational time needed to perform these simulations.

As in the case of a spherical black hole\cite{Anninos94c},
singularity avoiding slicings lead to large
gradients in metric functions that cannot presently be resolved in 3D and
eventually cause the code to crash.  The same problem occurs with
distorted black holes.  In Fig. \ref{fig:grr} we show the
radial metric function $\gamma_{rr}/\psi^{4}$, with its large round
peak, at time $t=27.2M$,
reconstructed from the Cartesian metric functions that are actually
evolved.  The spike developing near the origin is inside the throat,
and is a result of the application of the isometry condition.
One expects that the region near the metric function peaks needs to be
accurately computed in order to produce the correct waveform,
because the ringing radiation is produced by scattering off the Zerilli
potential,
which is located just outside of the peak as we know from studies of
horizon location\cite{Libson94a}. Although this potential is never explicitly
computed in the calculations, it is implicitly built into
the Cartesian metric functions being evolved.

\subsection{Radiation Extraction} Although in black hole simulations
we evolve directly the metric and extrinsic curvature, for
applications to gravitational wave astronomy we are particularly
interested in computing the waveforms emitted.  One measure of this
radiation is the Zerilli function, $\psi$, which is a gauge-invariant
function that obeys the Zerilli wave equation\cite{Zerilli70}.  The
Zerilli function can be computed by writing the metric as the sum
of a spherically symmetric part and a perturbation:
$g_{\alpha\beta}=\stackrel{o}{g}_{\alpha\beta}+h_{\alpha \beta}$,
where the
perturbation $h_{\alpha\beta}$ is expanded in tensor spherical harmonics.  To
compute the elements of $h_{\alpha\beta}$ in a numerical simulation,
one integrates the numerically evolved metric components
$g_{\alpha\beta}$ against appropriate
spherical harmonics over a
coordinate 2--sphere surrounding the black hole.  The resulting functions
can then be combined in
a gauge-invariant way, following the prescription given by
Moncrief\cite{Moncrief74}.  This procedure was originally developed by
Abrahams\cite{Abrahams88}, and was applied to the same class of
distorted black hole initial data sets discussed here, but evolved in
2D spherical--polar coordinates and with a different gauge, as
discussed in \cite{Abrahams92a}.

We have developed numerical methods based on the same ideas to extract
the waves in a full 3D Cartesian setting.  The method used is
essentially that used in the axisymmetric case, except that the metric
functions and their spatial derivatives need to be interpolated onto a
two-dimensional surface,
which we choose to have constant coordinate radius. The projections of
the perturbed metric functions $h_{\alpha\beta}$, and their radial
derivatives, are then computed by
numerically performing two-dimensional surface integrals for each
$\ell-m$ mode
desired.  Then, for each mode, the Zerilli function is constructed from
these projected metric functions, according to Moncrief's
gauge-invariant prescription.  This is a complicated but
straightforward procedure.  Both the numerical
interpolations and integrations involved in this extraction procedure
were chosen to be second
order accurate, and both have been shown to converge to second order
in the relevant grid spacing. As in Ref.~\cite{Abrahams92a}, we choose
to normalize the Zerilli function so that the asymptotic energy flux
in each mode is given by $\dot{E} = (1/32\pi) \dot{\psi}^2$.
While previously only axisymmetric
simulations have been studied, we can now study all non-trivial wave
modes, including those with $m\ne 0$.

We extracted the $\ell=2$ and $\ell=4$ Zerilli functions during an
evolution of the distorted black hole initial data set $(a,b,w,n,c) =
(0.5,0,1,2,0)$, using the extraction method described above.  In
Figure~\ref{fig:zer2dcomp}a we show the $\ell=2$ Zerilli function
extracted at a radius $r=8.7M$ as a function of time.  Superimposed on
this plot is the same function computed during the evolution of the
same initial data set with a 2D code, based on the one described in detail in
\cite{Abrahams92a,Bernstein93b}.  The agreement of the two plots
over the first peak is a strong affirmation of the 3D evolution code
and extraction routine.  It is important to note that the 2D results
were computed with a different slicing (maximal), different
coordinate system, and a {\em different spatial gauge}.  Yet the
physical results obtained by these
two different numerical codes, as measured by the waveforms, are
remarkably similar (as one would hope). This is the principal result
of this paper.  A full evolution with the
2D code to $t=100M$, by which time the hole has settled down to
Schwarzschild, shows that the energy emitted in this mode at that
time is about $4\times 10^{-3}M$.

In Fig.~\ref{fig:zer2dcomp}b we show the $\ell=4$ Zerilli function
extracted at the same radius, computed during evolutions with 2D and
3D codes. This waveform is more difficult to extract, because it has a
higher frequency in both its angular and radial dependence, and it has a
much lower amplitude: the energy emitted in this mode is three orders
of magnitude smaller than the energy emitted in the $\ell=2$ mode,
{\em i.e.}, $10^{-6}M$, yet it can still be accurately evolved and
extracted.

Small differences between the 2D and 3D results can be seen.
Resolution studies of the 3D results indicate that the differences are
not completely due to resolution of the 3D evolution code.  The small
differences in phase can be understood as a result of the different
shift and slicings being used in the two simulations.  The radiation is
extracted at a constant {\em coordinate} location, and the coordinates fall
towards the black hole at different rates with different slicings and
shifts.  By measuring the physical radial position of the wave
extraction in these simulations, we determined that the difference
between the 2D and 3D phases at late time is consistent with the
slightly different extraction locations in the two cases.  The
additional differences in the $\ell=4$ waveforms could be related to
slight differences in the initial data, which were generated in
independent ways, or even differences in gauge (the waveforms are
gauge-invariant, meaning they are unaffected only at first order under
gauge transformations).  As $\ell=4$ has a much smaller amplitude than
$\ell=2$, it will be more sensitive to such details.  The differences
are very small, and do not affect the conclusions of this paper, but
they will be studied in detail and discussed elsewhere.

\section{Summary and Conclusions}

We have shown that, in 3D numerical relativity, given sufficient
resolution, distorted black holes can be accurately evolved.
Furthermore
the gravitational waveforms generated by the black hole, consisting of
small perturbations on the
evolving black hole background, can be accurately propagated and
extracted from the numerically generated metric, on a 3D
Cartesian grid.  We have demonstrated this by comparing results from a
mature 2D code, showing good agreement not only for the $\ell=2$, but
also the $\ell=4$ modes of the radiation.

Although we regard this as an important step in establishing numerical
relativity as a viable tool to compute waveforms from black hole
interactions, the calculations one can presently do are limited.
With present techniques, the evolutions can only be carried out for a
fraction of the time required to simulate the 3D orbiting coalescence.
Many techniques to handle this more general case are under
development, such as hyperbolic formulations of the Einstein equations
and the advanced numerical methods they bring\cite{Bona97a}, adaptive
mesh refinement that will enable placing the outer boundary farther
away while resolving the strong field region where the waves are
generated, and apparent
horizon boundary conditions that excise the interiors of the black
holes, thus avoiding the difficulties associated with singularity
avoiding slicings.

All of these techniques, and others, may be needed to handle the more
general, long term evolution of coalescing black holes.  Our purpose
in this paper has been to show that {\em (a)} given present resources one
can evolve simpler distorted black hole systems and accurately extract the
waveforms, even when they carry only $10^{-6}M$ in energy, and {\em
(b)}
to establish testbeds for the techniques under development for the
more general case.  Each of these techniques may introduce numerical
artifacts, even if at very low amplitude, to which the waveforms may be
very sensitive.  As new methods are developed and applied to
numerical black hole simulations, they can now be tested on evolutions
such as those presented here to ensure that the waveforms are
accurately represented in the data.

In future papers we address the wave extraction in more detail;
work is presently in progress to apply it to more extensive
axisymmetric initial data \cite{Camarda97c}, to full 3D initial data
sets where nonaxisymmetric modes can be extracted for the first
time\cite{Camarda97a}, and to the evolution of colliding black holes
in 3D, extending the work in \cite{Anninos96c}.  Once this has been
fully developed and tested on full 3D data sets, it will be important
to apply it to true 3D black hole collision simulations, such as those
recently reported by Br{\"u}gmann \cite{Bruegmann97}.

\acknowledgements
This work has been supported by the Albert Einstein Institute (AEI)
and NCSA. We would like to thank K.V. Rao, John Shalf, and the staff at
NCSA for assistance with the computations.  Among many colleagues at
NCSA, AEI, and Washington University who have influenced this
particular work, we especially thank Andrew Abrahams, Gabrielle Allen,
Larry Smarr, and Wai-Mo Suen.  Calculations were performed at AEI and
NCSA on an SGI/Cray Origin 2000 supercomputer.



\begin{figure}
\caption{We show the radial metric function $\gamma_{rr}/\psi^{4}$ for
the evolution of the distorted black hole data set $(a,b,w,n,c)=(0.5,0,1,2,0)$
at time $t=27.2M$. The evolution was performed with $150^3$
grid points, although data from only the inner $100^3$ grid points are
shown to bring out detail. The resolution was $\Delta x=0.0544M$.}
\label{fig:grr}
\end{figure}

\begin{figure}
\caption{We show the (a) $\ell=2$ and (b) $\ell=4$ Zerilli functions
vs. time, extracted during 2D and 3D evolutions of the data set
$(a,b,w,n,c)=(0.5,0,1,2,0)$. The functions were extracted at a radius of
$8.7M$. The 2D data were obtained with $202\times 54$ grid points,
giving a resolution of $\Delta\eta=\Delta\theta=0.03$. The 3D data
were obtained using $300^3$ grid points and a resolution of $\Delta
x=0.0816M$.}
\label{fig:zer2dcomp}
\end{figure}

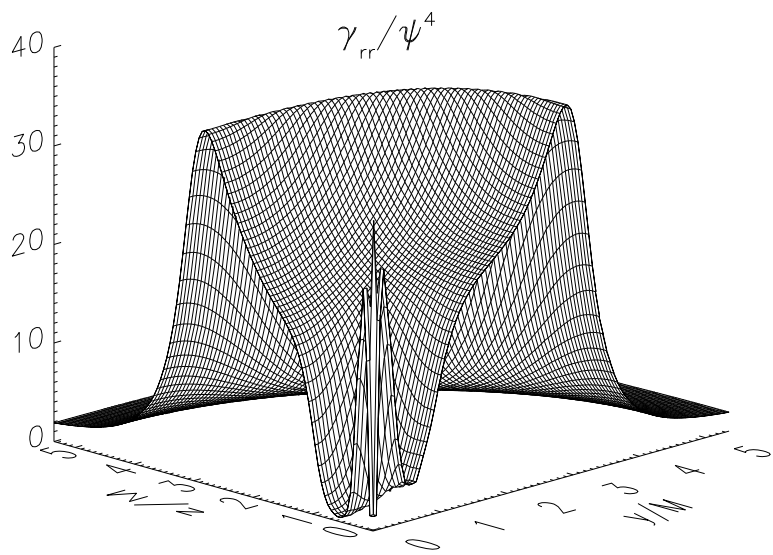
\end{document}